\documentclass[english,,showpacs,aps,prl,reprint,twocolumn]{revtex4-1}
\usepackage[T1]{fontenc}
\usepackage[latin9]{inputenc}
\usepackage{amssymb}
\usepackage{esint}
\usepackage{babel}

\begin{document}

\preprint{This line only printed with preprint option}

\title{Gapless Superfluid in $SU(2n+1)$ Fermions Systems}

\author{Betzalel Bazak}
\email{betzalel.bazak@mail.huji.ac.il}
\affiliation{The Racah Institute of Physics, The Hebrew University, 91904 Jerusalem,
Israel}
\author{Nir Barnea}
\affiliation{The Racah Institute of Physics, The Hebrew University, 91904 Jerusalem,
Israel}
\date{\today}

\begin{abstract}
We investigate the generalized Hubbard model of $(2n+1)$ Fermion species
interacting via a symmetric contact attraction potential.
We prove that the
ground state of such system is a gapless superfluid, where a full
Fermi surface coexists with a superfluid. Moreover, doing so we prove the existence of
a free mode in a strongly interacting system, regardless of the potential strength.
This proof holds
at the mean field level. A Grassmannian Gaussian integration
technique is used to deal with the problem. Our predictions may be
relevant to future high-spin cold atoms experiments.
\end{abstract}
\maketitle
The smooth crossover from a Bardeen-Cooper-Schrieffer (BCS) superfluidity,
valid for weakly interacting Fermions, to a Bose-Einstein Condensation
(BEC), valid for (composite) Bosons, was suggested theoretically about
30 years ago \cite{Leggett}. This prediction was verified experimentally
in the last decade by several groups using trapped cold atoms. Two-component
Fermionic atoms, like $^{40}\mathrm{K}$ \cite{K40} or $^{6}\mathrm{Li}$
\cite{Li6}, were trapped and the Feshbach resonance was used to control
the interaction between them. 

Recently, higher spin atoms were trapped, for which $F$, the hyper-fine
spin of the atom, enables studying systems of $(2F+1)$ Fermion species.
For example, an $F=1$ system was materialized with $\mathrm{^{6}Li}$ atoms
\cite{S1}, an $F=5/2$ with $^{173}\mathrm{Yb}$ atoms \cite{S5/2}, 
and $F=9/2$ with $^{87}\mathrm{Sr}$ atoms \cite{S9/2}. Note
that the experimental challenge in these systems is not limited
to trapping and cooling down
all the species of the relevant multiplet, but doing so while conserving
the $SU(N)$ symmetry. This goal is yet to be achieved in the BCS-BEC crossover
regime. This is due to the fact that the external magnetic field used
to control the interaction between the particles breaks the $SU(N)$
symmetry.

Experiments with $N=2$ cold atoms are used as a playground for condensed
matter physics \cite{ColdAtomRev}. Similarly, high-spin experiments
may shed light on other fields of physics. The $SU(3)$ case is relevant
for color-superconductivity in QCD, where the three quark colors generate
an $SU(3)$ symmetry. The $SU(4)$ case is relevant for nuclear physics,
where four types of Fermions exist (the proton and neutron with their
spin up or down), and form an approximate $SU(4)$ symmetry. It is well
known that pairing plays an important role in nuclear physics \cite{pairing}.
Therefore, the remarkable advance and flexibility of the cold atoms
experiments gives the opportunity of making experiments relevant to
other fields of physics which by themselves are hard to control and
measure. 

The generalization of superfluidity theory to the case of more
then two Fermion species arose soon after the BCS paper
was published \cite{BCS}, presenting the case of two band superconductivity
\cite{Suhl}. Since then, various techniques were used to deal with
the $SU(N)$ case, see for example Ref. \cite{SU(N)}. Considerable theoretical
attention was dedicated to the $SU(3)$ case \cite{SU(3),Gapless_su3,GapLess_HH}
and the $SU(4)$ case \cite{SU(4)}. Recently, guided by the experimental
development, the $SU(6)$ case was also studied \cite{SU(6)}.

A gapless superfluid, where a superfluid with gapped excitation spectrum
coexists with a gapless branch of excitation, is a well known phenomena
that naturally emerge in the case of particle symmetry breaking. For
example, when there is a difference in the density or the mass of
the interacting species, the result is a mismatch in their Fermi surfaces
and therefore a gapless superfluid emerges \cite{imbalance}. Here, in contrast,
we study a gapless superfluid in the symmetric case, where the $SU(N)$
symmetry is unbroken and completely holds. 

In this letter we focus on S-wave pairing in systems with odd number of species
$N=2n+1$. Starting with the $N=1$ case, 
i.e. one particle species, we note that pairing is forbidden and only
a gapless excitation can exist. For the $N=3$ case, it was proven
that a gapless excitation coexists with gapped superfluid excitation
\cite{Gapless_su3}. For the general case it was claimed but not proven,
that for any odd $N=2n+1$ the ground state of an $SU(N)$ symmetric
system is a gapless superfluid \cite{GapLess_HH}. Here we prove this
hypothesis to be true in general.

Let us consider a gas of $N$ species of Fermions, interacting via
a contact $SU(N)$-symmetric pairing potential. This system can be described
by the generalized Hubbard Hamiltonian,
\begin{eqnarray}
\lefteqn{H-\mu N =}\cr && \int
d{\bf
  x}\left[\sum_{\alpha}c_{\alpha}^{\dagger}\left(-\frac{\nabla^{2}}{2m}-\mu\right)c_{\alpha}
        -g\sum_{\alpha<\beta}c_{\alpha}^{\dagger}c_{\beta}^{\dagger}c_{\beta}c_{\alpha}
    \right],  
\end{eqnarray} 
where $c_{\alpha}^\dagger$ ($c_{\alpha}$) is the creation (annihilation)
operator for an $\alpha$  type  Fermion, $\mu$
is the chemical potential, and $g>0$ is the attractive contact pairing
coefficient. Note that the contact interaction 
coupling constant $g$ must be regularized. This is usually done through the
relation $m/4\pi a_s=g^{-1}+\sum_{k}(2\epsilon_{k})^{-1}$ between $g$ and
the scattering length $a_s$, which is the relevant physical parameter. 

Using the Hubbard-Stratonovich transformation \cite{HS} to decouple
the interaction in the Cooper channel, one gets the partition
function,
\begin{equation}
\mathcal{Z}=\int
  D(\bar{\Psi},\Psi)\int\prod_{\alpha<\beta}D(\bar{\Delta}_{\alpha\beta},\Delta_{\alpha\beta})e^{-S},
\end{equation}
where 
\begin{equation}\label{S}
S=\int_{0}^{\beta}d\tau\int
d^{3}x\left[\frac{1}{g}|\Delta|^{2}
    -(\bar{\Psi}\:\Psi)\mathcal{G}^{-1}\left(\begin{array}{l}  
\Psi^T \\
\bar\Psi^{T}\end{array}\right)\right].
\end{equation}
The field operator ${\Psi}=({\psi}_{1}\;{\psi}_{2}\;...{\psi}_{N})$
is a vector of coherent states, and $\Delta_{\alpha\beta}=-\Delta_{\beta\alpha}$
is the complex pairing field describing an S-wave Cooper pair consisting
of Fermions of types $\alpha$ and $\beta$. 
The total magnitude of the pairing fields is given by
$|\Delta|^{2}=\sum_{\alpha<\beta}|\Delta_{\alpha\beta}|^{2}$, 
 and $\mathcal{G}$ is
the extended Nambu-Gor'kov propagator. 

At this point let us assume that the pairing fields are fixed in time and
space $\Delta_{\alpha\beta}({\bf x},t)=\Delta_{\alpha\beta}$.
This simplifying assumption allows us to write 
$\mathcal{G}^{-1}$ in a block diagonal form in the momentum-frequency
$({\bf k},i\omega)$ space.
Each of these blocks is a $2N\times 2N$ matrix of the form
\begin{equation}\label{Gkw}
  \mathcal{G}^{-1}({\bf k},i\omega)=\left(\begin{array}{cc}
\openone(i\omega-\xi_{k}) & D\\
D^{\dagger} & \openone(i\omega+\xi_{k})\end{array}\right),\end{equation}
where $\openone$ is the $N\times N$ identity matrix, $\xi_{k}=k^{2}/2m-\mu$,
and $D$ is the pairing matrix $D_{\alpha\beta}=\Delta_{\alpha\beta}$. 
The corresponding $2N$ spinors are 
$(\bar\Psi\left( {\bf k}, i\omega),\Psi( -{\bf k}, -i\omega)\right)$, and the
sum on $\omega$ is limited to positive Matsubara frequencies only.

Note that the vector $(\bar{\Psi}\:\Psi)$ is not a standard Nambu spinor.
In order to consider all possible $N(N-1)$ pairing types 
the dimension of $\mathcal{G}^{-1}$ must be $2N\times2N$, and
therefore the spinors are $2N$-vectors. This raises the problem of
performing the integration over the $N$-vectors $\Psi,\bar{\Psi}$,
which in (\ref{S}) seems not to be Gaussian anymore. However in the block
diagonal form (\ref{Gkw}), the integration is
still Gaussian, and hence can be done quite easily. 

To show that, let us assume that $\left(\begin{array}{c}
\mathbf{u}_{i}\\
\mathbf{v}_{i}\end{array}\right)$ is an eigenvector of $\mathcal{G}^{-1}$ with the corresponding eigenvalue
$i\omega-\lambda_{i}$. One can easily verify that there exists a
twin eigenvector $\left(\begin{array}{c}
\mathbf{\mathbf{v}}_{i}^{*}\\
\mathbf{u}_{i}^{*}\end{array}\right)$ with the corresponding eigenvalue $i\omega+\lambda_{i}$. Hence,
the eigendecomposition of $\mathcal{G}^{-1}$ is given by $Q^{\dagger}\mathcal{G}^{-1}Q=\openone i\omega-\left(\begin{array}{cc}
\Lambda\\
 & -\Lambda\end{array}\right)$, where $Q=\left(\begin{array}{cccccc}
\mathbf{u}_{1} & ... & \mathbf{u}_{N} & \mathbf{v}_{1}^{*} & ... & \mathbf{v}_{N}^{*}\\
\mathbf{v}_{1} & ... & \mathbf{v}_{N} & \mathbf{u}_{1}^{*} & ... &
\mathbf{u}_{N}^{*}\end{array}\right)$,
 and $\Lambda$ is a real diagonal matrix composed of the eigenvalues
$\lambda_{i}$. Using this transformation the integration over the Fermionic
fields reads 
\begin{equation}
\int D(\bar{\Psi},\Psi)
e^{-(\bar{\Psi}\:\Psi)\mathcal{G}^{-1}\left(\begin{array}{l}
\Psi^T\\
\bar\Psi^{T}\end{array}\right)}=\prod_{i}(i\omega-\lambda_{i})(-i\omega-\lambda_{i}),
\end{equation}
where $D(\bar{\Psi},\Psi)$ stands for integration over the fields
$\Psi( {\bf k}, i\omega),\bar\Psi( {\bf k}, i\omega),
 \Psi(-{\bf k},-i\omega),\bar\Psi(-{\bf k},-i\omega)$. 
Restoring the momenta and frequencies summation one finally gets 
$\prod_{i,i\omega,\bf k}(i\omega-\lambda_{i})$.

Now that we have established the Grassmannian integration, we shall prove 
that for odd $N\geq 1$, there is always a gapless branch of
excitations, i.e. one of the eigenvalues of $\mathcal{G}^{-1}$ is
simply $i\omega-\xi_{k}$. To show that, we recall Jacobi's theorem
\cite{Jacobi} that if $M$ is a skew symmetric matrix of order $2n+1$
then the determinant of $M$ vanishes. 
Since $D^{\dagger}$ is such a matrix, one can conclude that there exists a vector $\tilde{\mathbf{v}}$
for which $D^{\dagger}\tilde{\mathbf{v}}=0$, therefore, 
$$
\left(\begin{array}{cc}
\openone(i\omega-\xi_{k}) & D\\
D^{\dagger} & \openone(i\omega+\xi_{k})\end{array}\right)\left(\begin{array}{c}
\tilde{\mathbf{v}}\\
0\end{array}\right)=(i\omega-\xi_{k})\left(\begin{array}{c}
\tilde{\mathbf{v}}\\
0\end{array}\right).$$
Consequently $\left(\begin{array}{c}
\tilde{\mathbf{v}}\\
0\end{array}\right)$ and $\left(\begin{array}{c}
0\\
\tilde{\mathbf{v}}^{*}\end{array}\right)$ are twin eigenvectors of $\mathcal{G}^{-1}$ with eigenvalues $i\omega-\xi_{k}$
and $i\omega+\xi_{k}$, respectively. These eigenvectors and eigenvalues correspond to a gapless
Fermi liquid excitation spectrum, thus proving our claim.

As an example, we explicitly consider the $SU(3)$ and $SU(5)$ cases.

In the case of three Fermion species, the gapless branch of excitations,
$\xi_{k}$, has the corresponding eigenvector $\left(\begin{array}{c}
\tilde{\mathbf{v}}\\
0\end{array}\right)$, where $\tilde{\mathbf{v}}=\left(\begin{array}{r}
\bar{\Delta}_{23}\\
-\bar{\Delta}{}_{13}\\
\bar{\Delta}_{12}\end{array}\right)$ and $D^{\dagger}\tilde{\mathbf{v}}=0$. The spectra of the two gapped
branches are $E_{k}=\pm\sqrt{\xi_{k}^{2}+|\Delta|^{2}}.$ The mean
field values of $\Delta_{\alpha\beta}$ can be found from the saddle
point approximation $\partial S/\partial\bar{\Delta}_{\alpha\beta}=0$
yielding three identical gap equations,
\begin{equation}
\frac{1}{g}=\sum_{k}\frac{1}{2E_{k}}\tanh\frac{\beta E_{k}}{2}.
\end{equation}
Note that the same gap equation is valid for the $SU(2)$ case, resulting
in the same magnitude of the (total) pairing gap and the same critical
temperature. The density equation for the $\alpha$ Fermion species
is,
\begin{equation}
  \rho_{\alpha}=\sum_{k}\left[f_{\alpha}n_{F}(\xi_{k})+(1-f_{\alpha})
               \frac{1}{2}\left(1-\frac{\xi_{k}}{E_{k}}\tanh\frac{\beta      
    E_{k}}{2}\right)\right],\label{eq:density}
\end{equation}
where $f_{\alpha}=|\Delta_{\beta\gamma}|^{2}/|\Delta|^{2}$. If the
system is balanced, i.e. contains the same number of particles for each kind,
we get $|\Delta_{12}|=|\Delta_{23}|=|\Delta_{13}|$ and $f_{\alpha}=1/3$.
That is to say that one third of the system remains gapless while
the residual two thirds are superfluid. Note that since the different
$\Delta_{\alpha\beta}$'s are equal, none of the species is unpaired,
but instead there is a collective gapless excitation.

In the case of five Fermion species, the gapless branch of excitations 
has the corresponding eigenvector $\left(\begin{array}{c}
\tilde{\mathbf{v}}\\
0\end{array}\right)$, where 
$\tilde{\mathbf{v}}=\left(\begin{array}{r}
\bar{\eta}_{1}\\
-\bar{\eta}_{2}\\
\bar{\eta}_{3}\\
-\bar{\eta}_{4}\\
\bar{\eta}_{5}\end{array}\right)$, 
$$\eta_{\alpha}=\Delta_{\beta\gamma}\Delta_{\delta\epsilon}-\Delta_{\beta\delta}\Delta_{\gamma\epsilon}+\Delta_{\beta\epsilon}\Delta_{\gamma\delta},$$
and $D^{\dagger}\tilde{\mathbf{v}}=0$. The spectra of the four gapped
branches are 
$E_{k}=\pm\sqrt{\xi_{k}^{2}+\frac{1}{2}|\Delta|^{2}\pm\frac{1}{2}\sqrt{|\Delta|^{4}-4|\eta|^{2}}}$,
where $|\eta|^{2}=\sum_{\alpha}|\eta_{\alpha}|^{2}.$ Note that the
cases where $|\Delta|^{4}=4|\eta|^{2}$ and $|\eta|^{2}=0$ will be
of specific interest, as the former has two doubly degenerated gapped
branches and the latter has three gapless branches and two gapped
branches. 

To conclude, we have studied a system of $N$-Fermion species with an 
attractive, on-site, $SU(N)$ symmetric interaction. We have proved that for odd $N$,
the ground state of the system is a gapless superfluid.
Doing so, we have used a generalized Grassmannian Gaussian
integration. The key point in our proof is the fact that the determinant of
the skew-symmetric pairing matrix vanishes if its dimension is odd. As a
result a gapless free Fermi mode appears. 
The cases $N=3$ and $N=5$ were solved analytically at the mean field level 
to demonstrate our general claim. We hope that future cold atoms experiments 
would confirm our predictions.

\section*{Acknowledgments}
This work was supported by the ISRAEL SCIENCE FOUNDATION 
(Grant No.~954/09). The authors would like to thank 
D. Gazit, D. Orgad and Y. Abarbanel for their useful
comments and suggestions in the course of preparing this work.

\end{document}